\documentstyle[multicol,eqsecnum,aps,epsf]{revtex}

\draft

\begin{document}
\title{How Far Can the SO(10) Two Higgs Model \\[.2in]
Describe the Observed Neutrino Masses and Mixings ? }

\author{\bf K.~Matsuda, Y.~Koide$^{(a)}$, T.~Fukuyama, 
and H. Nishiura$^{(b)}$}
\address{%
Department of Physics, Ritsumeikan University, 
Kusatsu, Shiga, 525-8577 Japan \\
(a) Department of Physics, University of Shizuoka, Shizuoka 422-8526 
Japan\\
(b) Department of General Education, Junior College of Osaka Institute 
of Technology, \\
        Asahi-ku, Osaka 535-8585, Japan}%
\date{17 August 2000}

\maketitle
\begin{abstract}
Can the SO(10) model with one {\bf 10} and one {\bf 126} Higgs scalars 
give the observed masses and mixings of quarks and leptons without
any other additional Higgs scalars?
Recently, at least, for quarks and charged leptons, it has been 
demonstrated that it is possible. However, for the neutrinos, 
it is usually said that parameters which are determined from the quark 
and charged lepton masses cannot give the observed large neutrino 
mixings. This problem is systematically investigated, and it is
concluded that the present data cannot exclude SO(10) model with
two Higgs scalars although it cannot give the best fit values
of the data.
\end{abstract}

\begin{multicols}{2}
\section{Introduction}

SO(10) GUT model seems to us the most attractive model when we take
the unification of the quarks and leptons into consideration.
However, in order to reproduce the observed quark and lepton masses 
and mixings, usually, a lot of Higgs scalars 
are brought into the model.  So it is the very crucial problem 
to know the minimum number 
of the Higgs scalars which can give the observed fermion mass spectra 
and mixings.  A model with one Higgs scalar is obviously ruled out 
for the description of the realistic quark and lepton mass spectra. 
Two Higgs models were initially discussed by Mohapatra et.al \cite{mohapatra}.

In the previous paper \cite{matsuda}, we discussed 2 Higgs scalars, 
\{{\bf 10} and {\bf 126}\} case and \{{\bf 10} and {\bf 120}\} case, 
and showed that they reproduce quark-lepton mass matrices unlike 
the conventional results \cite{takasugi}.  One of new points of 
our approach is that we adopt general forms of Yukawa couplings 
allowable in the SO(10) framework.
However, we did not argue there about the neutrino mass matrix 
since it may incorporate additional assumptions like the seesaw 
mechanism etc.  
\par
One of the merits of the SO(10) model is that it includes a right-handed 
Majorana neutrinos in the fundamental representation and naturally 
leads to the seesaw mechanism. Also some papers claim 
that two Higgs model (\{{\bf 10} and {\bf 126}+$\overline{{\bf 126}}$\}) 
does not reproduce the large mixing angle of the atmospheric neutrino 
deficit \cite{brahmachari}.
So in this paper we apply our method developed in the previous paper 
to the neutrino mass matrix, fitting the other parameters of the 
quark-lepton mass matrices.
Our model has the two Higgs scalars \{{\bf 10} and {\bf 126}\}
 both of which are symmetric with respect to the family index.  
Therefore mass matrices are symmetric whose entries are complex valued.
We do not adopt another choice \{{\bf 10} and {\bf 120}\}.  
For it does not involve the mass term of the right-handed Majorana 
neutrinos which are the ingredients of the seesaw mechanism.
\par
We begin with the short review of our previous work \cite{matsuda}.
In the case where two Higgs scalars, $\phi_{10}$ and $\phi_{126}$, are 
incorporated in the SO(10) model, the mass matrices of quarks and charged 
leptons have the following forms
\begin{equation}
M_u=c_0M_0+c_1M_1,~ M_d=M_0+M_1,~ M_e=M_0- 3 M_1. 
\label{M1}
\end{equation}
Here $M_0$ and $M_1$ are the mass matrices generated by the Higgs scalars 
$\phi_{10}$ and $\phi_{126}$, respectively. Also $c_0$ and $c_1$ are 
the ratios of VEV's,
\begin{eqnarray}
c_0&=&v_0^u/v_0^d=\langle\phi^{u0}_{10}\rangle/\langle\phi^{d0}_{10}\rangle,
\nonumber \\
c_1&=&v_1^u/v_1^d=\langle\phi^{u0}_{126}\rangle/\langle\phi^{d0}_{126}\rangle, 
\label{c0c1}
\end{eqnarray}
and $\phi^u$ and $\phi^d$ denote Higgs scalar components which couple with 
up- and down-quarks, respectively. 
Eliminating $M_0$ and $M_1$ from Eq.(\ref{M1}), we obtain
\begin{equation}
M_e=c_dM_d+c_uM_u, \label{M2}
\end{equation}
where
\begin{equation}
c_d=-
{\frac{3c_0+c_1}{c_0-c_1}} \ ,\ \ 
c_u={\frac{4}{c_0-c_1}} \ .
\end{equation}
Since $M_u$, $M_d$, and $M_e$ are complex symmetric matrices, they are 
diagonalized
by unitary matrices $U_u$, $U_d$, and $U_e$, respectively, as 
\begin{equation}
U_u^{T}M_uU_u=D_u \ , \ \
U_d^{T}M_dU_d=D_d \ , \ \
U_e^{T}M_eU_e=D_e \ , \label{diag1}
\end{equation}
where $D_u$, $D_d$, and $D_e$ are diagonal matrices given by
\begin{eqnarray}
&&D_u \equiv \mbox{diag}(m_u,m_c,m_t) \ , \ \
  D_d \equiv \mbox{diag}(m_d,m_s,m_b) \ , \nonumber \\
&&D_e \equiv \mbox{diag}(m_e,m_\mu,m_\tau) \ , \label{diag2}
\end{eqnarray}
Since the 
Cabibbo-Kobayashi-Maskawa (CKM) matrix $V_q$ is given by 
\begin{equation}
V_q=U_u^{T} U_d^* \ ,
\end{equation}
the relation (\ref{M2}) is re-written as follows: 
\begin{equation}
(U_e^{\dagger}U_u)^T D_e(U_e^{\dagger}U_u)
=c_{d}V_q D_{d}V_q^{T}
+c_uD_u .\label{eq071703}
\end{equation}
Therefore, we obtain the independent three equations:
\end{multicols}
\hspace{-0.5cm}
\rule{8.7cm}{0.1mm}\rule{0.1mm}{2mm}
\widetext
\begin{eqnarray}
{\rm Tr}D_e D_e^\dagger &=& |c_d|^2 \, {\rm Tr}
\Bigl[(V_q D_{d}V_q^{T}+\kappa D_u)
(V_q D_{d}V_q^{T}+\kappa D_u)^\dagger\Bigr],
 \label{eq82501}\\
{\rm Tr}(D_e D_e^\dagger)^2 &=&|c_d|^4 \,
{\rm Tr}\Bigl[
((V_q D_{d}V_q^{T}+\kappa D_u)
(V_q D_{d}V_q^{T}+\kappa D_u)^\dagger)^2\Bigr],
 \label{eq82502}\\
{\rm det}D_eD_e^\dagger &=& |c_d|^6 \, {\rm det}
\Bigl[(V_q D_{d}V_q^{T}+\kappa D_u)
(V_q D_{d}V_q^{T}+\kappa D_u)^\dagger\Bigr],
\label{eq82503}
\end{eqnarray}
\hspace{9.2cm}
\rule[-2mm]{0.1mm}{2mm}\rule{8.7cm}{0.1mm}
\begin{multicols}{2}
\narrowtext
\noindent
where $\kappa=c_u/c_d$. 
By eliminating the parameter \(c_d\), we have two equations 
for the parameter \(\kappa\):
\begin{eqnarray}
\frac{(m_e^2+m_\mu^2+m_\tau^2)^3}{m_e^2 m_\mu^2 m_\tau^2}&=&
  \frac{(\ref{eq82501})^3}{(\ref{eq82503})}, \label{eq82511}\\
\frac{(m_e^2+m_\mu^2+m_\tau^2)^2}
     {2(m_e^2 m_\mu^2+m_\mu^2 m_\tau^2+m_\tau^2 m_e^2)}&=&
\frac{(\ref{eq82501})^2}{(\ref{eq82501})^2-(\ref{eq82502})}, 
\label{eq82512}
\end{eqnarray}
where \((\ref{eq82501})^3\), for instance, 
means the right-hand side of Eq.(\ref{eq82501}) to the third power.
Let us denote the parameter values of $\kappa$ evaluated from 
Eqs.(\ref{eq82511}) and (\ref{eq82512}) as ${\kappa}_A$ and ${\kappa}_B$, respectively.  
If ${\kappa}_A$ and ${\kappa}_B$ coincide with each other, 
then we have a possibility 
that the SO(10) GUT model can reproduce the observed quark and charged 
lepton mass spectra.  
If ${\kappa}_A$ and ${\kappa}_B$ do not so, the SO(10) model with 
one {\bf 10} and one {\bf 126} Higgs scalars is ruled out, 
and we must bring 
more Higgs scalars into the model.  
\par
Note that Eqs. (\ref{eq82501})-(\ref{eq82503}) can constrain only 
the absolute value of \(c_d\equiv |c_d|e^{i\sigma}\). 
The argument of the parameter \(c_d\) can be determined 
by taking neutrino sector 
into consideration.
In the previous paper \cite{matsuda}, we have found that only for
the signs of the masses 
\begin{eqnarray}
&&(m_t, m_c, m_u; m_b, m_s, m_d;  m_\tau, m_\mu,
m_e)\nonumber\\
&& \hspace{2.3cm}=(+,-,+; +,-,-; +,\pm, \pm)~~\mbox{(a)},\\ 
&&\mbox{and}\nonumber\\
&& \hspace{2.3cm}=(+,-,-; +,-,-; +,\pm, \pm)~~ \mbox{(b)}, 
\end{eqnarray}
there are solutions which gives $\kappa_A=\kappa_B$, and the corresponding
parameter values $(|c_d|, \kappa)$ are
\begin{eqnarray}
(|c_d|, \kappa)
&=&(3.15698, -0.019296e^{2.64172^\circ i}),\\
&& (3.03577, -0.019398e^{2.99570^\circ i})~~\mbox{for (a)},\\
\mbox{and}\hspace{0.5cm}&&\nonumber\\
&=&(3.13307, -0.019314e^{2.71464^\circ i}), \label{data1}\\
&& (3.00558, -0.019420e^{3.10014^\circ i})~~\mbox{for (b)}
\end{eqnarray}
and $m_s=76.3$ [MeV] for input $\theta_{23}=0.0420$ [rad] and 
$\delta =60^\circ$ at $\mu=m_Z$ ($m_Z$ is the neutral weak boson mass).
For the relation between the values at $\mu=m_Z$ and those
at $\mu=\Lambda_X$ ($\Lambda_X$ is a unification scale), see
Ref.\cite{matsuda}.
The purpose of the present paper is to investigate whether these
solutions can give reasonable values for observed neutrino masses 
and mixings or not.

\section{The number of parameters in the SO(10) model with two Higgs scalars}
As we have discussed in the previous section, 
among four freedoms of complex $\{c_0,~c_1\}$ or $\{c_d ,~\kappa\}$, 
we have been able to fix the three of them, $\kappa$ and $|c_d|$.  
This is not accidental.
Let us discuss the situation in detail in the SO(10) two Higgs model. 

In the previous paper, by using the relation (\ref{eq071703}),
we have investigated whether there is a set of parameters which
can give the 13 observable quantities $D_e$, $D_u$, $D_d$, and $V_q$
or not.
We can rewrite Eq.(\ref{eq071703}) as
\begin{equation}
A_e^T D_eA_e=c_d (V_q D_{d}V_q^{T}+\kappa D_u), \label{eq01080701}
\end{equation}
where 
\begin{equation}
A_e = U_e^\dagger U_u,
\end{equation}
\begin{equation}
c_d = |c_d| e^{i\sigma} .
\end{equation}
The quantities $D_e$, $D_u$, $D_d$, and $V_q$
are inputs, and the quantities $|c_d|$, $\kappa$, and $A_e$ are
the parameters which should be fixed from those observed quantities.
In general, an $n\times n$ unitary matrix for \(n\) generations has $n^2$ parameters.
Therefore, the number of the parameters is
\begin{equation}
N(\mbox{pmt}) = N(A_e) +N(c_d)+N(\kappa) = n^2+2+2.
\end{equation}
On the other hand, the number of equations is
\begin{equation}
N(\mbox{eqs}) = n(n+1) ,
\end{equation}
because Eq.(\ref{eq01080701}) is symmetric.
Therefore, the number of the unfixed parameters is given by
\begin{equation}
N_{{\rm free}}= N(\mbox{pmt}) -N(\mbox{eqs}) = 4-n = 1 ,
\end{equation}
for $n=3$, i.e., the 13 observed quantities fix the parameters
$|c_d|$, $\kappa$, and $A_e$, but 1 parameter \(\sigma\) 
remains as an unknown parameter.

In the present paper, we will try to predict neutrino masses
\begin{equation}
D_\nu = U_\nu^T M_\nu U_\nu ,
\end{equation}
and mixing matrix
\begin{equation}
V_\ell = U_e^T U_\nu^* ,
\end{equation}
by using the observed quantities $D_e$, $D_u$, $D_d$, and $V_q$
and the parameter values $|c_d|$, $\kappa$, and $A_e$ fixed by
Eq.(\ref{eq01080701}).

SO(10) GUT asserts that the Dirac neutrino mass matrix $M_D$
is given by the form 
\begin{equation}
M_D = c_0 M_0- 3 c_1 M_1,
\end{equation}
and 
Majorana mass matrices of the left-handed and right-handed neutrinos,
$M_L$ and $M_R$, are proportional to the matrix $M_1$:
\begin{equation}
M_L = c_L M_1, \ \ \ M_R = c_R  M_1,
\end{equation}
where $M_0$ and $M_1$ are related to the quark and charged lepton
mass matrices $M_u$, $M_d$, and $M_e$ as follows:
\begin{eqnarray}
M_0&=&\frac{3M_d+ M_e}{4},\label{eq081401}\\
M_1&=&\frac{M_d-M_e}{4}.\label{eq071701}
\end{eqnarray}
Then the neutrino mass matrix derived form the seesaw mechanism becomes
\begin{eqnarray}
M_\nu&=&M_L-M_DM_R^{-1}M_D^T\nonumber\\
 &=&c_LM_1 \nonumber\\
 & &-c_R^{-1}(c_0M_0-3c_1M_1)M_1^{-1}(c_0M_0-3c_1M_1)^T. 
\label{eq071702}
\end{eqnarray}
In the present paper we adopt $c_L=0$.
Also we may ignore the phase of \(c_R\) 
which does not affect the observed values.
Therefore, we can rewrite Eq.(\ref{eq071702}) as
\begin{equation}
|c_R| A_\nu^T D_\nu A_\nu = \widetilde{M}_D \widetilde{M}_1^{-1}
                            \widetilde{M}_D^T, \label{eq081004}
\end{equation}
similarly to Eq.(\ref{eq01080701}), where
\begin{eqnarray}
\widetilde{M}_D&=&c_0 \widetilde{M}_0- 3 c_1 \widetilde{M}_1,\\
\widetilde{M}_0&=&\frac{1}{4} (3\widetilde{M}_d+ \widetilde{M}_e ),\\
\widetilde{M}_1&=&\frac{1}{4} ( \widetilde{M}_d-\widetilde{M}_e ), 
\label{eq0717xx}
\end{eqnarray}
with
\begin{eqnarray}
\widetilde{M}_d &=& U_u^T M_d U_u = V_q D_d V_q^T , \\
\widetilde{M}_e &=& U_u^T M_e U_u = A_e^T D_e A_e 
\nonumber \\
                &=& c_d ( V_q D_d V_q^T +\kappa D_u).
\end{eqnarray}
Differently from the previous work, 
the quantities $D_\nu$ and
$V_\ell$ are unknown parameters at the present stage.
Since
\begin{equation}
V_\ell = A_e^* A_\nu^T,\label{eq081201}
\end{equation}
and $A_e$ is fixed from Eq.(\ref{eq01080701}), 
the number of the unknown parameters in Eq.(\ref{eq081201}) is
\begin{equation}
N(A_\nu) =N(V_\ell) = n^2.
\end{equation}
Of course, the unknown parameters in \(A_\nu\) contain 
the \(n\) unphysical parameters 
which cannot be determined because of the rephasing in the fields $e_L$.
Therefore, the number of the unknown parameters is
\begin{eqnarray}
N(\mbox{pmt}) & = & N(D_\nu)+N(A_\nu)+N(|c_R|)+N(\sigma) \nonumber \\
 & = & n+n^2+1+1=n^2+n+2
\end{eqnarray}
and from the number of equations 
$N(\mbox{eqs})=n(n+1)$ in Eq.(\ref{eq081004}),
we obtain the number of the unfixed parameters as
\begin{eqnarray}
N_{{\rm free}}&=&N(\mbox{pmt})-N(\mbox{eqs}) \nonumber \\
  &=& (n^2+n+2) -n(n+1)= 2.
\end{eqnarray}
This means that we can predict neutrino masses and mixing
completely if we give the two values $|c_R|$ and $\sigma$.
The numerical predictions will be investigated in the next
section.

\section{Numerical Results}

Here we discuss the numerical results of the neutrino mass spectrum 
and neutrino mass matrix.
For example, we use the set in Eq.(\ref{data1}). 
Even if the other sets are used, our results are scarcely changed.
The allowed values of neutrino mass square differences and lepton 
flavor mixing angles depict complicated tracks with moving 
$\sigma \equiv \arg c_d$ (Fig. 2).
This figure shows a general tendency that the lepton flavor mixing angles 
\(\theta_{12}\) and \(\theta_{23}\) get larger as $\sigma$ approaches to $3\pi/2$.
For an illustration we take \(\sigma=149\pi/100\), then these values become
\begin{eqnarray}
&&\frac{\Delta m_{12}^2}{\Delta m_{13}^2} = 0.15, \hspace{0.8cm}
\frac{\Delta m_{23}^2}{\Delta m_{13}^2} = 0.85, \nonumber\\
&&\sin^2 (2\theta_{12}) = 0.76, \quad \sin^2 (2\theta_{23}) = 0.75,
\nonumber\\ 
&&\sin^2 (2\theta_{13}) = 0.16.
\label{fitting1}
\end{eqnarray}
There still remain a little bit discrepancies between our results and 
experiments.
However our results are much improved in comparison with those by 
Babu-Mohapatra \cite{mohapatra} in which they obtained  
\(\sin\theta_{12}=0-0.3\), \(\sin\theta_{13}=0.05\), and 
\(\sin\theta_{23}=0.12-0.16\).
The purpose of the present paper is to study the general tendency of 
the fittings and not to pursuit the precise data fitting,   
for the data themselves are not affirmative and we have 
theoretical ambiguities not incorporated in the present data fitting 
like the renormalization group effect.

In the choice of Eq.(\ref{fitting1}), we have
\begin{eqnarray}
|c_d| &=& 3.16\\
c_0 &=&  \frac{1 - c_d}{c_u} = 54.84 e^{-20.24^\circ i}, \\
c_1 &=& -\frac{3 + c_d}{c_u} = 70.54 e^{+41.90^\circ i}.
\end{eqnarray}
In this case, Eq.(\ref{eq081401}) - Eq.(\ref{eq071702}) are re-written 
in the basis of $M_u=D_u$ (see Eq.(\ref{eq071703})) as
\end{multicols}
\widetext
\begin{eqnarray}
M_0 &=& \frac{3V_q D_d V_q^T
          + c_d(\kappa D_u + V_q D_d V_q^T)}{4}\nonumber \\
    &=& 2.1646\times 10^3 e^{+10.48^\circ i} \left(
        \begin{array}{lll}
	-0.00405 e^{-57.29^\circ i} & -0.00753 e^{-56.24^\circ i} &
        -0.00533 e^{+65.46^\circ i} \\
	-0.00753 e^{-56.24^\circ i} & -0.02986 e^{-51.59^\circ i} & 
        +0.06358 e^{-57.64^\circ i} \\
	-0.00533 e^{+65.46^\circ i} & +0.06358 e^{-57.64^\circ i} &
        +1.00000
        \end{array}
        \right) \mbox{[MeV]}, \\
M_1 &=& \frac{V_q D_d V_q^T
         - c_d(\kappa D_u + V_q D_d V_q^T)}{4} \nonumber\\
    & =& 9.5127\times 10^2 e^{-24.44^\circ i} \left(
        \begin{array}{lll}
	-0.00715 e^{+95.23^\circ i} & -0.01333 e^{+96.54^\circ i} &
        +0.00944 e^{+38.23^\circ i} \\
	-0.01333 e^{+96.54^\circ i} & -0.04878 e^{+90.73^\circ i} &
        +0.11247 e^{+95.13^\circ i} \\
	+0.00944 e^{+38.23^\circ i} & +0.11247 e^{+95.13^\circ i} &
        +1.00000
        \end{array}
        \right)\mbox{[MeV]},\\
|c_R| M_\nu &=& (c_0 M_0-3 c_1 M_1) M_1^{-1} (c_0 M_0-3 c_1 M_1)^T
\nonumber \\
    & =& -4.6628 \times 10^6 e^{-52.17^\circ i} \left(
        \begin{array}{lll}
	+0.1163 e^{+26.89^\circ i} &  +0.2165 e^{+28.06^\circ i} &
        -0.1536 e^{-30.53^\circ i} \\
	+0.2165 e^{+28.06^\circ i} &  +0.8193 e^{+28.00^\circ i} &
        -1.9276 e^{+29.52^\circ i} \\
	-0.1536 e^{-30.53^\circ i} &  -1.9276 e^{+29.52^\circ i} &
        +1.0000
        \end{array}
        \right)\mbox{[MeV]}. \label{eq01080702}
\end{eqnarray} 
Let us choose the free parameter \( |c_R| \) so as to result in small 
neutrino masses, for example when \( |c_R|=3.2\times 10^{14}\), 
we have \(\Delta m_{23}^2=1.5\times 10^{-3}\)[eV\(^2\)].

Here there arises a question what makes the two flavor mixing 
angles large.
We need to investigate 
the mixing matrices \(U_e\) and \(U_\nu\) which diagonalize \(M_e\) and \(M_\nu\), respectively.
Those are obtained as
\begin{eqnarray}
U_e &=& \left(
	\begin{array}{lll}
	+0.863                         & +0.504 e^{+\ 9.46^\circ i} &
        -0.022 e^{+ 56.66^\circ i} \\
	-0.493 e^{-\ 9.82^\circ i}     & +0.834          &
        -0.248 e^{+ 16.63^\circ i} \\
	-0.110 e^{- 21.40^\circ i}     & +0.223 e^{- 18.10^\circ i} &
        +0.969
        \end{array}
        \right),\\
%
U_\nu &=& \left(
	\begin{array}{lll}
	+0.992                      &  -0.092 e^{-15.94^\circ i} &
        -0.088 e^{+ 12.86^\circ i} \\ 
	+0.049 e^{+ 76.86^\circ i}   &  +0.724    &
        -0.688 e^{- 16.08^\circ i} \\ 
	+0.117 e^{+\ 9.80^\circ i}   &  +0.683 e^{+16.74^\circ i} &
        +0.721     
        \end{array}
        \right).
\end{eqnarray}
Here, \(|U_{e11}|\), \(|U_{e12}|\), \(|U_{e21}|\), \(|U_{e22}| \agt 0.5\) 
for 
the charged lepton mass matrix and 
\(|U_{\nu22}|\), \(|U_{\nu23}|\), \(|U_{\nu32}|\), \(|U_{\nu33}| \agt 0.7\) 
for
the neutrino mass matrix.
Therefore the components of the lepton flavor mixing matrix become
\(|V_{l11}|\), \(|V_{l12}|\), 
\(|V_{l21}|\), \(|V_{l22}|\), 
\(|V_{l23}|\), \(|V_{l32}|\), 
\(|V_{l33}| \agt 0.5\):
\begin{equation}
V_{l} = \left(
	\begin{array}{ccc}
	+0.844 e^{+\ 2.10^\circ i}  & -0.494 e^{-\ 9.95^\circ i} &
        +0.206 e^{+ 23.61^\circ i} \\
	+0.527 e^{+\ 3.26^\circ i}  & +0.696 e^{-\ 8.84^\circ i} &
        -0.488 e^{+ 24.97^\circ i} \\ 
	+0.098 e^{- 15.78^\circ i}  & +0.521 e^{- 27.43^\circ i} &
        +0.848 e^{+\ 6.32^\circ i}
        \end{array}
        \right).
\end{equation}
The mixing angle \(\theta_{23}\) becomes larger, 
while the mixing angle \(\theta_{12}\) smaller,
if we take the smaller value of \(|m_t|\) or \(|m_d|\), 
or larger \(|m_c|\), or \(|m_b|\), \(|m_s|\) than their center values.

As a simple example, 
the shift of \(|m_d|\) and \(|m_s|\) causes 
the change of mixing angles and neutrino mass square 
differences as depicted in Fig.2.
Fig.2 shows that the \(\theta_{23}\) and \(\theta_{13}\) can approach 
the \(99\%\)C.L. of SK\cite{SK00}+CHOOZ\cite{CH00} but 
\(\theta_{12}\) and \(\Delta m_{12}^2\) are out of the range of
\(99\%-99.9\%\)C.L. of SOLAR\cite{Nu2000}+CHOOZ.

\begin{figure}[bhp]
\begin{center}
\leavevmode
\epsfxsize=17.0cm
\epsfbox{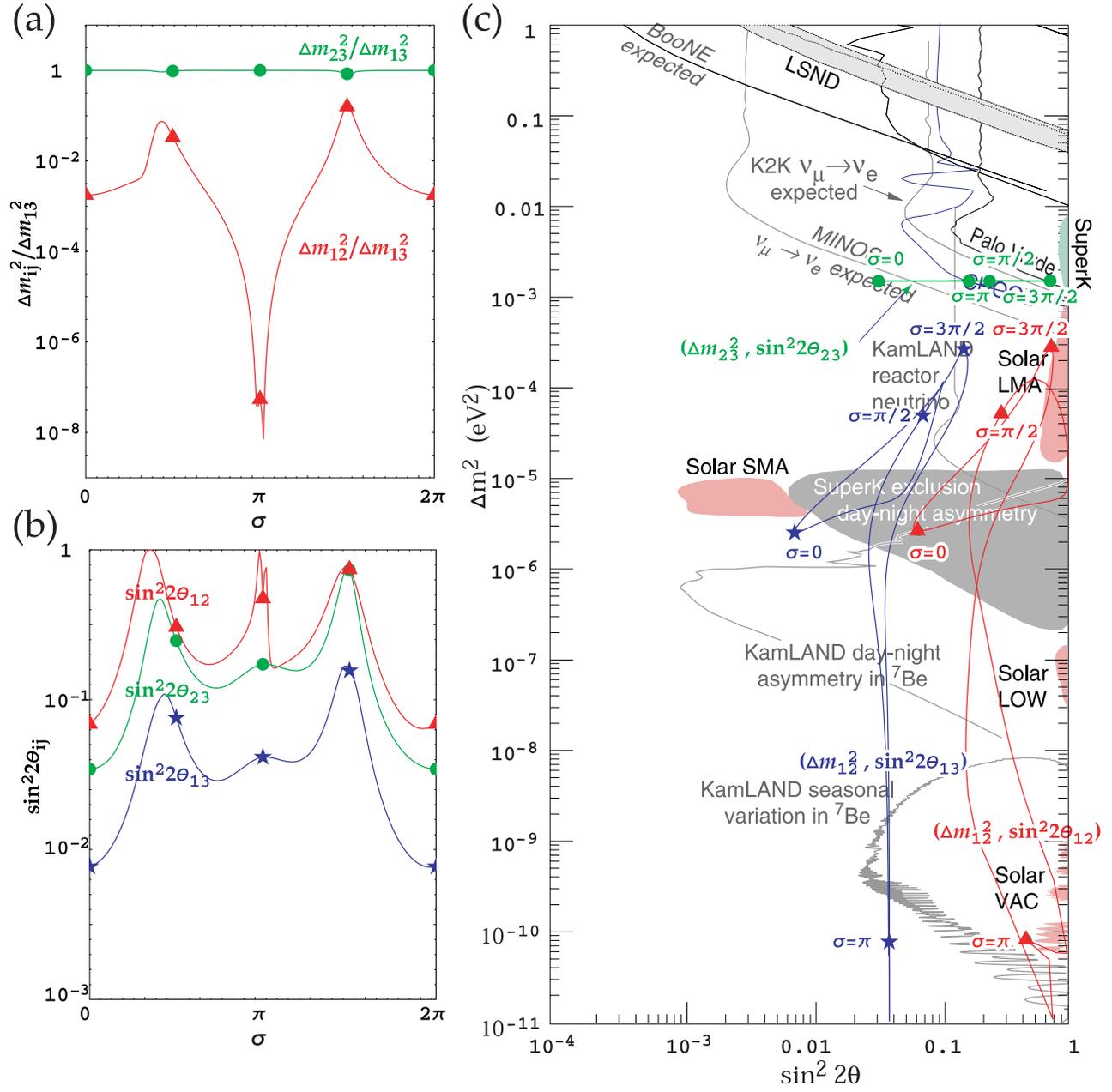}
\end{center}
\caption{
The relation between our results and the
two-flavor oscillation analysis [6] 
when $\sigma$ is moved.
(a) The circles and triangles indicate the values of
$\Delta m_{23}^2/\Delta m_{13}^2$ and $\Delta m_{12}^2/\Delta m_{13}^2$ 
at every $\pi/2$ of $\sigma$.
(b) The circles, triangles, and stars indicate the values of
$\sin^2 2\theta_{23}$,
$\sin^2 2\theta_{12}$, and $\sin^2 2\theta_{13}$
at every $\pi/2$ of $\sigma$, respectively.
(c) The circles, triangles, and stars indicate the values of
$(\Delta m_{23}^2, \sin^2 2\theta_{23})$, 
$(\Delta m_{12}^2, \sin^2 2\theta_{12})$, and
$(\Delta m_{12}^2, \sin^2 2\theta_{13})$ 
at every $\pi/2$ of $\sigma$. 
Here we have set $\Delta m_{23}^2=1.5\times 10^{-3}[\mbox{eV}^2]$
in every case.}
\label{fig2}
\end{figure}

\begin{figure}[bhp]
\begin{center}
\leavevmode
\epsfxsize=12.75cm
\epsfbox{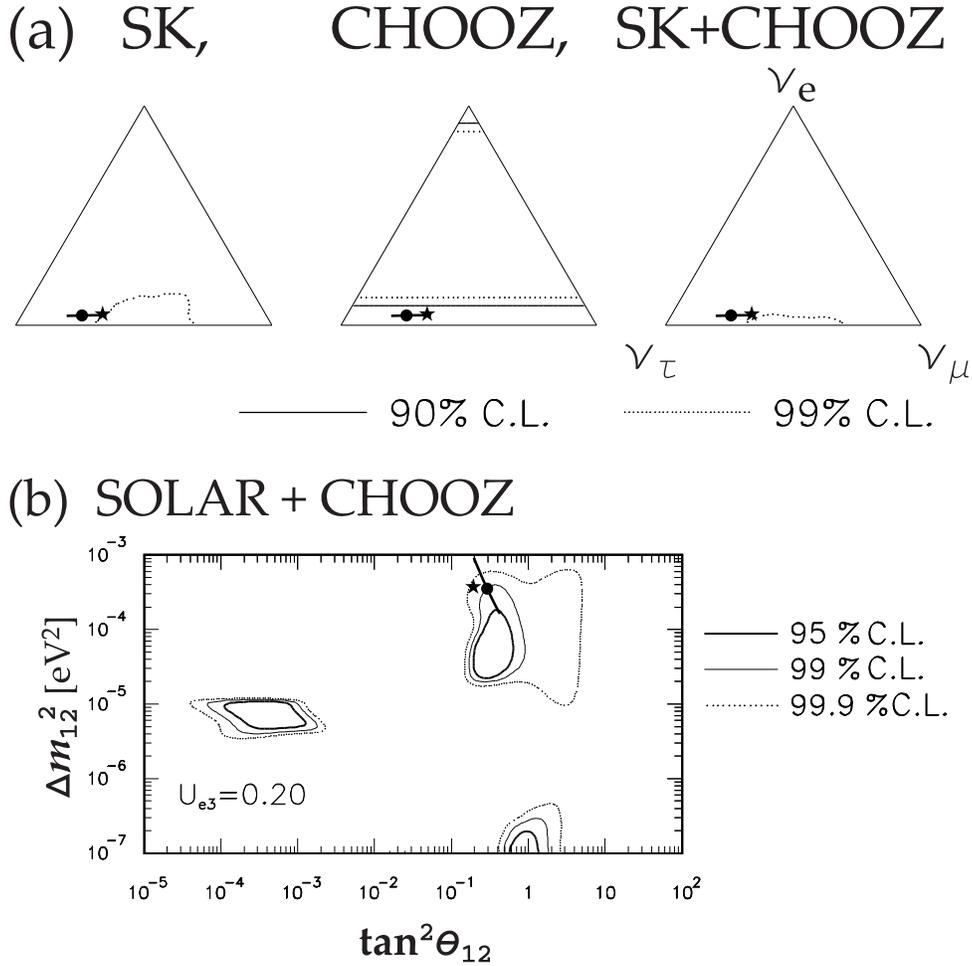}
\end{center}
\caption{
The relation between 3$\nu$ Oscillation analyses by G.L.Fogli et al. [10] 
and by us for $\Delta m_{23}^2=1.5\times 10^{-3}
[\mbox{eV}^2] $.
(a) For SK+CHOOZ. (b) For SOLAR+CHOOZ.
The circles indicate our solutions for Eq.(\ref{fitting1}). 
The solid line through them is the track as $m_d$ is varied.
It goes from the experimental limit that  $|m_d|$ moves over the range, 
4.03 - 5.29 {\rm MeV} [5].
At that time,  $|m_s|$ simultaneously changes over the range, 76.3 - 76.2 {\rm MeV} 
so as to satisfy the relations (\ref{eq82511}) and (\ref{eq82512}).
If we take the smaller $|m_d|$ with the fixed $\sigma$, the solution 
in (a) moves rightward and the solution in (b) does left-upward (Table I (i)).
Since the minimum $|m_d|$ for (b) gives bad fitting, 
we have changed $\sigma$ from $149\pi/100$ to $146\pi/100$ ,
which is denoted by star (Table I (ii)).
Thus our result approaches the $99\%$C.L. of SK+CHOOZ and $99.9\%$C.L.
of SOLAR+CHOOZ.} 
\label{fig3}
\end{figure}
\hspace{9.2cm}
\begin{multicols}{2}
\narrowtext
\section{Discussion}
Since there are only two basic matrix \(M_0\) and \(M_1\) in this model, 
the number of parameters in Eq.(\ref{eq01080701}) and (\ref{eq081004}) is
\begin{equation}
\begin{tabular}{lr}
 \(D_u\), \(D_d\), \(D_e\), \(D_\nu\) & 3\(\times\)4 = 12\\
 \(c_d\), \(|c_R|\), \(\kappa\)       & 2+1+2 = \hspace{0.8mm} 5\\
 \(V_q\), \(A_e\), \(A_\nu\)          & 4+9+9 = 22\\
\hline
 sum.                                 & 39
\end{tabular}
\end{equation}
and the number of equations is \(N(\mbox{eqs})=12 \times 2 = 24\).
Therefore the number of free parameters is \(N(\mbox{pmt})-N(\mbox{eqs})=39-24=15\).
On the other hand, the number of the physical parameters 
which can be determined by experiments is 
\begin{equation}
\begin{tabular}{lr}
  \(m_u\), \(m_c\), \(m_t\),          & 3\\
  \(m_d\), \(m_s\), \(m_b\),          & 3\\
  CKM: \(\theta_{12}\), \(\theta_{23}\), 
       \(\theta_{13}\), \(\delta\),   & 4\\ 
  \(m_e\), \(m_\mu\), \(m_\tau\),     & 3\\
  \(m_{\nu_e}\), \(m_{\nu_\mu}\), 
  \(m_{\nu_\tau}\),                          & 3\\
  MNS: \(\theta_{12}\), \(\theta_{23}\), 
       \(\theta_{13}\),
      \(\delta\), \(\beta\), \(\rho\) & 6\\
\hline
 sum.                                 & 22
\end{tabular}
\end{equation}
where \(\beta\) and \(\rho\) are Majorana phases in the Maki-Nakagawa-Sakata (MNS) matrix
because of no rephasing in the neutrino fields $\nu_L$.
To sum up the matter, we discuss the consistency test about 22 physical parameters 
by using only 15 free parameters.
The consistency test in the quark sector is good, as shown in our previous paper.
In the lepton sector, the test is not so bad 
when we adopt the MSW large mixing angle solution of solar neutrino deficit, 
and this model favors the normal hierarchy of neutrino mass spectrum.

Also we can predict the yet unobserved values 
such as the averaged neutrino masses \(\langle m \rangle_{\alpha\beta}\) 
and Jarlskog parameter in the lepton part.
The averaged neutrino masses appear in the reactions where the Majorana neutrinos 
propagate in the intermediate states. They are 
\begin{equation}
\langle m_\nu\rangle_{\alpha\beta}\equiv \left|
\sum_{j=1}^3U_{\alpha j}U_{\beta j}m_j \right|,
\end{equation}
where \(\alpha\) and \(\beta\) are $(e,~\mu,~\tau)$.  
They correspond to neutrinoless double beta decay \cite{takasugi2} 
for $\alpha=\beta=e$, $\mu -e$ conversion 
($\mu^-+(A,Z)\rightarrow e^++(A,Z-2)$ for $\alpha=\mu,~\beta=e$, 
and \(K\) decay ($K^-\rightarrow \pi^+\mu^-\mu^-$) for 
$\alpha=\beta=\mu$ \cite{fukuyama} etc.
In Fig.3 we have depicted $\sigma$ dependence of 
$\langle m_\nu\rangle_{\alpha\beta}/\sqrt{\Delta m_{23}^2}$.
In the case of Eq.(\ref{fitting1}), these values become as follows.
\begin{eqnarray}
&&\frac{\langle m \rangle_{\alpha\beta}}{\sqrt{\Delta m_{23}^2}}
\simeq 
\left(
\begin{array}{ccc}
0.032 & 0.065 & 0.30 \\
      & 0.096 & 0.59 \\
      &       & 0.67
\end{array}
\right).
\end{eqnarray}
For instance, if we input \(\Delta m_{23}^2=1.5 \times 10^{-3}\) [eV\(^2\)],
\(\langle m \rangle_{ee}\) becomes 0.0012[eV].
This is smaller than the experimental value of the next generation experiments 
such as GENIUS\cite{genius}, CUORE\cite{cuore}, and MOON\cite{moon}.
Jarlskog parameter \cite{jarlskog} appears in three generations
\begin{eqnarray}
P(\nu_e \rightarrow \nu_\mu)-P(\nu_\mu \rightarrow \nu_e)=
J\frac{\Delta E_{21}\Delta E_{32}\Delta E_{31}}{\Delta E^M_{21}
\Delta E^M_{32}\Delta E^M_{31}}\nonumber\\
\times\mbox{sin}\left( \frac{\Delta E^M_{21}L}{2}\right)
\mbox{sin} \left( \frac{\Delta E^M_{32}L}{2} \right )
\mbox{sin}\left( \frac{\Delta E^M_{31}L}{2}\right)
\end{eqnarray}
with
\begin{equation}
J \equiv \mbox{Im}(V_{l12} V_{l22}^* V_{l13}^* V_{l23}).
\end{equation}
Here we have adopted the notation
\begin{eqnarray}
\Delta E_{jk}&\equiv& E_j-E_k=\frac{\Delta m_{jk}^2}{2E}\nonumber\\
\Delta E^M_{jk}&\equiv& E^M_j-E^M_k
\end{eqnarray}
with
\begin{eqnarray}
U\mbox{diag}(E_1,E_2,E_3)U^{-1}+\mbox{diag}(a,0,0)\nonumber\\
\equiv U^M\mbox{diag}(E^M_1,E^M_2,E^M_3)(U^M)^{-1}
\end{eqnarray}
The $\sigma$ dependence of $J$ is depicted in Fig.4. 
For Eq.(\ref{fitting1}), it takes
\begin{equation}
J \simeq 0.00015.
\end{equation}
However, it needs careful consideration that 
\(J\) drastically changes at \(\sigma \simeq 3\pi/2\).  
\(\langle m \rangle_{\alpha\beta}\) and \(J\) 
in the cases of Table I (i) and (ii) discussed in Fig.2  
are also listed in Table II (i) and (ii).
In this paper we have discussed 
how far the SO(10) two Higgs scalar model describes the quark-lepton masses and 
mixing parameters. 
We can conclude that this model cannot be rejected within the existing data. 
It should be remarked that 
the whole parameters can be decided from the existing data in principle.

\section*{acknowledgement}
\vspace{-2cm}
The work of K.M. is supported by the JSPS Research Grant No. 10421.

\vspace{8cm}
\begin{figure}[htbp]
\begin{center}
\leavevmode
\epsfxsize=8.5cm
\epsfbox{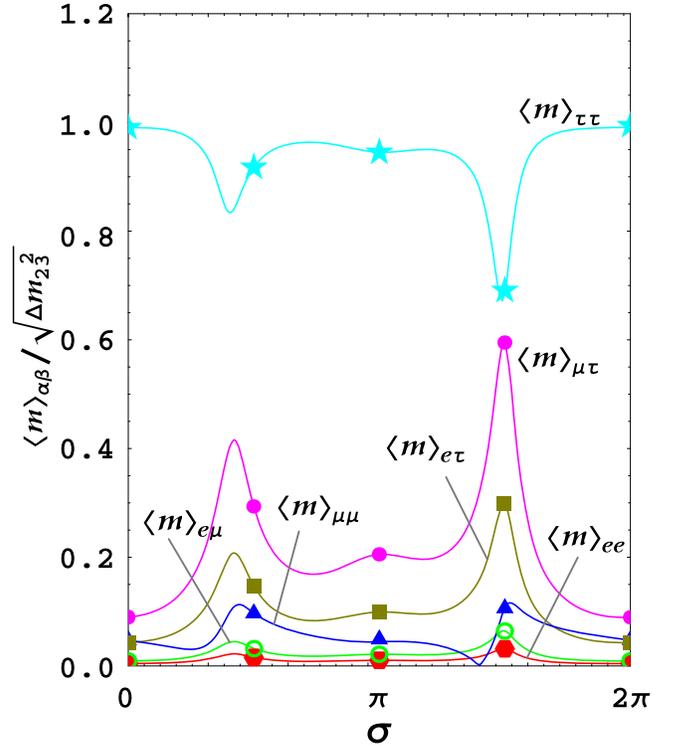}
\end{center}
\caption{
The relations between the averaged neutrino masses of 
lepton number violation process and \(\sigma\). 
The hexagons, white circles, boxes, triangles, 
black circles, and stars indicate the values of
\(\langle m \rangle_{ee      }/\sqrt{\Delta m_{23}^2}\), 
\(\langle m \rangle_{e\mu    }/\) \(\sqrt{\Delta m_{23}^2}\), 
\(\langle m \rangle_{e\tau   }/\sqrt{\Delta m_{23}^2}\), 
\(\langle m \rangle_{\mu\mu  }/\sqrt{\Delta m_{23}^2}\), 
\(\langle m \rangle_{\mu\tau }/\sqrt{\Delta m_{23}^2}\), and 
\(\langle m \rangle_{\tau\tau}/\sqrt{\Delta m_{23}^2}\) 
at every $\pi/2$ of $\sigma$, respectively.
}
\label{fig4}
\end{figure}

\begin{figure}[bhp]
\begin{center}
\leavevmode
\epsfxsize=8.5cm
\epsfbox{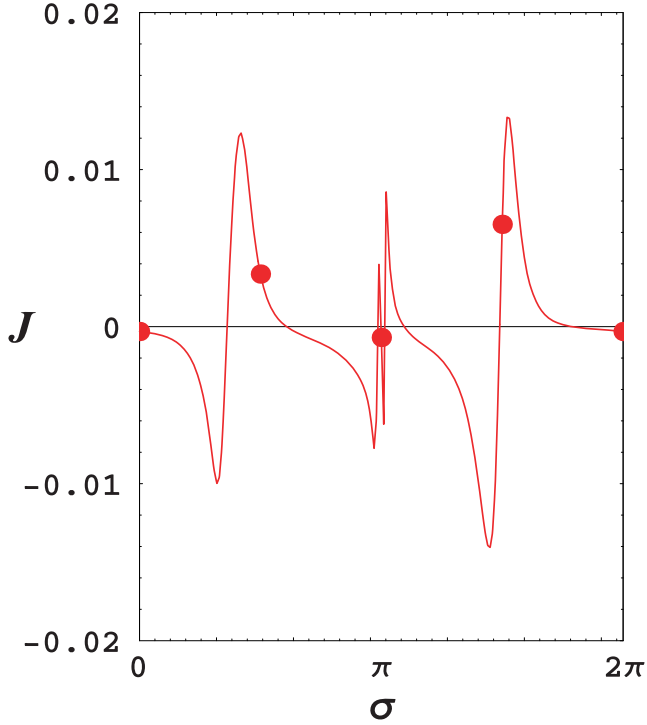}
\end{center}
\caption{
The relation between 
Jarlskog parameter $J$ and \(\sigma\).
The circles indicate the values of \(J\)
at every $\pi/2$ of $\sigma$.
}
\label{fig5}
\end{figure}

\narrowtext

\begin{table}[htbp]
\begin{tabular}{c|l}
(i) & \(|m_d|=4.03\)[MeV],\hspace{0.1cm} 
      \(|m_s|=76.3\)[MeV],\hspace{0.1cm} \(\sigma=149\pi/100\), \\
    & \((\Delta m_{12}^2)/(\Delta m_{13}^2)=0.43\), \hspace{0.55cm}
      \((\Delta m_{23}^2)/(\Delta m_{13}^2)=0.57\),\\
    & \(\sin^2(2\theta_{12})=0.52\), \(\sin^2(2\theta_{23})=0.91\), 
      \(\sin^2(2\theta_{13})=0.17\)\\ \hline
(ii) & \(|m_d|=4.03\)[MeV],\hspace{0.1cm} 
      \(|m_s|=76.3\)[MeV],\hspace{0.1cm} \(\sigma=146\pi/100\), \\
    & \((\Delta m_{12}^2)/(\Delta m_{13}^2)=0.20\),  \hspace{0.55cm}
      \((\Delta m_{23}^2)/(\Delta m_{13}^2)=0.80\),\\
    & \(\sin^2(2\theta_{12})=0.54\), \(\sin^2(2\theta_{23})=0.88\), 
      \(\sin^2(2\theta_{13})=0.20\)
\end{tabular}
\caption{
Our solution (the second and third lines) from the input parameters (the first line).
The result of (i) is obtained 
when we move \(|m_d|\) from 4.69[MeV] to 4.03[MeV].
(ii) is the result when we move \(|m_d|\) as (i) 
and, furthermore, \(\sigma\) from \(149\pi/100\) to \(146\pi/100\).
These data fitting corresponds to Fig.2.
}
\end{table}

\begin{table}[htbp]
\begin{tabular}{c|l}
(i) & \(\langle m \rangle_{ee      }/\sqrt{\Delta m_{23}^2}=0.039\), \qquad
      \(\langle m \rangle_{e\mu    }/\sqrt{\Delta m_{23}^2}=0.086\), \\
    & \(\langle m \rangle_{e\tau   }/\sqrt{\Delta m_{23}^2}=0.43\hspace{1.3mm} \), \qquad
      \(\langle m \rangle_{\mu\mu  }/\sqrt{\Delta m_{23}^2}=0.19\hspace{1.3mm} \), \\
    & \(\langle m \rangle_{\mu\tau }/\sqrt{\Delta m_{23}^2}=0.96\hspace{1.3mm} \), \qquad
      \(\langle m \rangle_{\tau\tau}/\sqrt{\Delta m_{23}^2}=0.52\hspace{1.3mm} \), \\
    & \(J=0.0091\)\\ \hline
(ii)& \(\langle m \rangle_{ee      }/\sqrt{\Delta m_{23}^2}=0.028\), \qquad
      \(\langle m \rangle_{e\mu    }/\sqrt{\Delta m_{23}^2}=0.064\), \\
    & \(\langle m \rangle_{e\tau   }/\sqrt{\Delta m_{23}^2}=0.32\hspace{1.3mm} \), \qquad
      \(\langle m \rangle_{\mu\mu  }/\sqrt{\Delta m_{23}^2}=0.10\hspace{1.3mm} \), \\
    & \(\langle m \rangle_{\mu\tau }/\sqrt{\Delta m_{23}^2}=0.71\hspace{1.3mm} \), \qquad
      \(\langle m \rangle_{\tau\tau}/\sqrt{\Delta m_{23}^2}=0.54\hspace{1.3mm} \), \\
    & \(J=-0.014\)
\end{tabular}
\caption{
The values of averaged neutrino masses and 
the Jarlskog parameter for the case (i) and (ii) in Table I.
}
\end{table}

\end{multicols}

\end{document}